# Expert exploranation for communicating scientific methods - A case study in conflict research

Benedikt Mayer[*,†]      Karsten Donnay[‡]      Kai Lawonn[§]
Bernhard Preim[†]      Monique Meuschke[†]

## Preprint Version

### Abstract

Science communication aims at making key research insights accessible to the broad public. If explanatory and exploratory visualization techniques are combined to do so, the approach is also referred to as *exploranation*. In this context, the audience is usually not required to have domain expertise. However, we show that exploranation can not only support the communication between researchers and a broad audience, but also between researchers directly. With the goal of communicating an existing method for conducting causal inference on spatio-temporal conflict event data, we investigated how to perform exploranation for experts, i.e., *expert exploranation*. Based on application scenarios of the inference method, we developed three versions of an interactive visual story to explain the method to conflict researchers. We abstracted the corresponding design process and evaluated the stories both with experts who were unfamiliar with the explained method and experts who were already familiar with it. The positive and extensive feedback from the evaluation shows that expert exploranation is a promising direction for visual storytelling, as it can help to improve scientific outreach, methodological understanding, and accessibility for researchers new to a field.

---

[*]Corresponding author: benedikt@isg.cs.uni-magdeburg.de

[†]Department of Simulation and Graphics, Otto-von-Guericke-University Magdeburg, Universitätsplatz 2, Magdeburg, 39106, Germany

[‡]Department of Political Science & Digital Society Initiative, University of Zurich, Affolternstrasse 56, Zurich, 8050, Switzerland

[§]Institute for Informatics, Friedrich-Schiller-Universität Jena, Ernst-Abbe-Platz 2, Jena, 07743, Germany

# 1 Introduction

When communicating research results that are based on real data to a broad audience, it is helpful to make use of *visual storytelling*. As "visual language spans borders of knowledge, experience, age, gender, and culture" (Ynnerman, Löwgren & Tibell, 2018, p. 13), it is well-suited to support the explanation of technical concepts in an understandable way. To make the process even more efficient, conveying the information in the form of a *visual story* is beneficial (Gershon & Page, 2001). The engagement and understanding of the users of such visual stories can be increased further by allowing them to interactively explore aspects of the conveyed information on their own Ynnerman, Löwgren & Tibell (2018). Yet, the exploration should happen in a controlled way to prevent users from getting lost in the data. This combination of explanation and exploration is referred to as *exploranation* (Ynnerman, Löwgren & Tibell, 2018). It is commonly used to convey information from domain experts to a broad audience without domain expertise (Ynnerman, Löwgren & Tibell, 2018). However, we argue that it is also beneficial for supporting the communication between experts from the same domain, particularly for explaining the workings of scientific methods.

The scenario we focus on is the following: Researchers have developed a complex method for which it is not straightforward to assess how and under which circumstances it can be applied. It could support the work of fellow researchers, but they do not have the capacity to delve deeply enough into the method to determine whether it would be beneficial to them. In this scenario, exploranation can facilitate the fellow researchers to gain insights into the workings of the method to assess its applicability for their own research problems.

So far, not a lot of research has been performed in this direction, while there exist several questions regarding such exploranation for domain experts, i.e., *expert exploranation* (ExEx). The questions include whether researchers are interested in ExEx overall, how they would use it, and which potential directions they see for it. The process and challenges of creating ExEx stories need to be investigated. This includes what kind of information is necessary for other experts to understand the workings of a complex method in sufficient detail, how to abstract this information, and how to order it in an understandable sequence. Moreover it is of interest how to perform effective storyboarding for producing suitable visualizations and text using appropriate jargon.

To investigate the process and potential of ExEx in more detail, and in which aspects it differs from performing exploranation for broad audiences, we cooperated with a conflict researcher to create a story to explain one of their developed methods (Schutte & Donnay, 2014). The method's purpose is to perform counterfactual inference between spatio-temporal conflict events. For instance, it was used to test which kinds of aid projects are suitable to stabilize war-torn communities, and which, instead, tend to increase anti-government activities (Karell & Schutte, 2018). We considered three such application scenarios to automatically generate three versions of the same story to explain the causal inference method (Mayer, 2023). This way, we aim at increasing the insights into how ExEx can be performed in the following way:

- We present the design process of an interactive story to explain a conflict research method to domain experts.

- We present the results of an evaluation of the story. The evaluation was performed with two complementary user groups – one with in-depth knowledge of the explained method and one with barely any prior knowledge.

- We summarize our learnings regarding how ExEx can be performed and which future research is necessary.

## 2 Related work

We structure this section based on science communication, in general, and data-driven visual storytelling.

### 2.1 Science communication

**Audience.** To explain research insights to a target audience, first, the audience needs to be understood (Meredith, 2021). According to Meredith, there is a finer distinction than only between an expert and a lay broad audience (Meredith, 2021). We focus on conflict researchers, ranging from graduate students to senior scientists. As even scientists can struggle with numeracy, i.e., the ability to understand numeric information (National Academies of Sciences Engineering and Medicine, 2017), it is recommendable to reduce the cognitive effort, e.g., by explaining what certain numbers mean, and drawing attention to important information (Institute of Medicine, 2014). How members of an audience interpret new information also depends on their beliefs, values, and ways of understanding the world (National Academies of Sciences Engineering and Medicine, 2017). The corresponding "mental models" (Bruine de Bruin & Bostrom, 2013) can be expected to better support a correct understanding of conveyed concepts if an audience member has domain expertise (National Academies of Sciences Engineering and Medicine, 2017).

**Visualization.** Meredith advocates using informative and engaging visuals for communicating scientific information to improve the comprehension and engagement of the audience (Meredith, 2021). There are many examples for communicating science to broad audiences by making use of visualization, e.g., for astro physics (Lan et al., 2021), climate communication (Windhager, Schreder & Mayr, 2019), and medical communication (Meuschke et al., 2022). However, for addressing expert audiences, the literature is more sparse. The most common means for communicating within the research community are traditionally abstracts, posters, oral presentations, peer-reviewed research papers, and, more recently, open access formats (Udovicich, Kasivisvanathan & Winchester, 2017). Not a lot of focus is put on using effective and engaging ways of communication. It can even be considered unprofessional to present the content "too entertainingly."

**Storytelling.** Yet, writing scientific articles in a more enjoyable and interesting way does not automatically reduce the information gained from them (Hunsaker, 1979). Moreover, employing techniques from traditional storytelling to communicate science can even improve the memorization of the conveyed information while making it more enjoyable (Negrete & Lartigue, 2010). The combination of science and stories can be supported by using appropriate story arcs (Green, Grorud-Colvert & Mannix, 2018) and incorporating data visualization (Ma et al., 2012), thereby using techniques from visual storytelling (Tong et al., 2018).

## 2.2 Data-driven visual storytelling

Visual storytelling can be used to tell any kind of story while incorporating visualization, not just scientific ones. If the story and its visualizations depend on data, as in our case, it is also referred to as *data-driven visual storytelling* (Riche et al., 2018).

The following works provide guidelines on how to tell such stories.

**Story creation process.** For the general steps of our story creation process, we primarily referred to five works (Lee et al., 2015; Cortes Arevalo et al., 2020; Zhang et al., 2022; Satyanarayan & Heer, 2014; Amini et al., 2015). Lee et al. (2015) identified steps ranging from the initial ideation, over the scripting and editing, to the final presentation to the audience. Contrary to our work, they did not focus on a specific audience. More closely related in that regard is the work of Cortes Arevalo et al. (2020), presenting key steps to take when creating stories for the communication between researchers and practitioners. Zhang et al. (2022) focused on cognitive and communication theories in their framework for data-driven visual storytelling. They adapted their design process from Satyanarayan & Heer (2014). Both groups of authors (Zhang et al., 2022; Satyanarayan & Heer, 2014), as well as Lee et al. (2015), assume the design process of a visual data story to begin with the exploration of a data set in search for interesting facts to communicate.

In contrast, in our scenario, the key goal of the story is already known, i.e., explaining a scientific method. In this case, the beginning of the design process is less a task of exploring data, but more a task of identifying which core mechanics of the method should be explained in what detail. These considerations are part of the larger process of storyboarding, which was analyzed by Amini et al. (2015), revealing the non-linear and iterative order of the steps involved. In our work, we present the similarities and differences between our expert-focused story creation process and the five works listed above.

**Story structure and layout.** Our story follows the traditional Freytag's Pyramid story arc, for which Yang et al. (2021) analyzed how it can be applied to data stories. Regarding the overall layout, our story would be classified as a "dynamic slideshow" (Roth, 2021) or "interactive slideshow" (Segel & Heer, 2010), as it is navigated discretely through a set of slides and provides interactions at key points throughout the narrative (Segel & Heer, 2010). Hullman et al. (2013) investigated how to sequence slideshow-style presentations for improving their comprehensibility and memorability. Following their advice, we tried to minimize the transition costs between different views of our story.

**Storytelling techniques.** For advice regarding concrete storytelling techniques, we referred to the design space for spatio-temporal data stories created by Mayer et al. (2023), and the three design spaces are based on Roth (2021), Stolper et al. (2016)

and Mayr & Windhager (2018). Such design spaces often lead to the development of authoring tools, based on the identified dimensions. However, in their survey of authoring tools from 2023, Chen et al. (2023) came to the conclusion that tools to create complex slideshows are still lacking. Overall, there is still a gap between fully automated story generation and manually created stories (Sun et al., 2023).

**Structuring visualization and text.** Previous works inform on how visualizations and text can be integrated effectively (Zhi, Ottley & Metoyer, 2019), some even focusing on spatio-temporal stories (Latif, Chen & Beck, 2021). Taken together with McKenna et al. (2017) narrative *flow factors*, including aspects like how the reader can navigate through the story, these works provide a foundation for structuring and improving the story reading experience.

**Engagement and interaction.** Engaging users is particularly helpful when communicating complex topics (Böttinger et al., 2020). Factors influencing user engagement include a story's aesthetics, whether it can spark the users' interest, and how well it can keep their attention (Hung & Parsons, 2017). At that, users can go through cycles of engagement and disengagement (O'Brien & Toms, 2008). To reengage users, it can help to give them control over the story to discover new insights (Hung & Parsons, 2017), e.g., by allowing them to interactively explore aspects of the story. Hohman et al. (2020) extensively discuss how online articles can use interaction to engage users, potentially improving learning outcomes. The work itself is an interactive article published on *Distill*.[1] This website, along with others, like explorable explanations,[2] provide a variety of examples for articles communicating scientific content in an interactive and engaging way. When avoiding the pitfall of hiding important information behind interaction, interactive exploration can be used to dig deeper into the data and build trust with the story (Aisch, 2023). While user engagement is an even more central aspect when communicating to a broad audience, experts can also benefit from it for the reasons listed above.

Coining the term *exploranation*, Ynnerman, Löwgren & Tibell (2018) present design principles for how to combine interactive exploration with explanation. They recommend augmenting explanative applications like data stories with constrained explorative microenvironments. In applications of the exploranation approach, it was claimed that corresponding systems can be used for lay audiences as well as expert audiences, only by adjusting the underlying story (Bock et al., 2018; Brossier et al., 2023). Similarly, we also show that generating multiple versions of the same base story with differing content and level of detail can be beneficial. However, in contrast to the two works (Bock et al., 2018; Brossier et al., 2023), we argue that dedicated investigations focusing explicitly on experts are still necessary. In the two works, the focus was put on contextualizing and exploring scientific data sets while using the same underlying system, *OpenSpace* (Bock et al., 2018). However, instead of patterns in a data set, we aim at explaining the workings of a scientific method, for which several concepts need to be introduced, building on each other. This requires customized sequences of conceptual explanations and corresponding views. Applications based on *OpenSpace* that go in such a more concept-oriented direction tend to do so in a live presenter setting. However, our goal was to investigate how to create stories that do not need a live presenter but can be consumed asynchronously in a self-contained manner. Accordingly, the requirements for our story differ from those of the settings based on *OpenSpace*, targeting broad audiences.

---

[1] https://distill.pub/; accessed 23 October 2023
[2] https://explorabl.es/; accessed 23 October 2023

Another related approach are *interactive lecture demonstrations*, where lectures are enriched by allowing students to get more actively involved with the presented information (Sokoloff & Thornton, 2004). However, this approach is also typically applied in synchronous settings with a live presenter. In summary, we use insights from science communication to create a visual data story for communicating the workings of a scientific method between researchers. For the design, we build on a broad range of works providing guidelines for the creation of slideshow-based, interactively explorable stories. With our work, we do not intend to replace traditional means for communicating scientific results between researchers (Udovicich, Kasivisvanathan & Winchester, 2017), but to enrich the existing set of options with a more accessible and immediate way to learn about new scientific methods.

# 3    The communicated method

We give a brief outline of the method that we introduce in our expert exploranation, called *matched wake analysis* (MWA) (Schutte & Donnay, 2014). To do so, we follow the example of how MWA was applied in an analysis performed in a separate paper (Karell & Schutte, 2018). Below, we refer to this example application as *EA*. The EA study focuses on Afghanistan to analyze whether aid projects that exclude parts of a given community can lead to an increase in anti-government (i.e., *insurgent*) activities in that area. An example of an aid project excluding part of a community would be funding a selected private group of residents from a larger community. An example of an insurgent activity would be a violent attack by a terrorist group against the military or civilians supporting the government. The goal of the study was to not only identify the correlation between aid projects and insurgent violence, but rather to determine whether the exclusionary application of aid *causes* an increase in insurgent attacks. To answer this question, the authors made use of MWA to perform a corresponding causal inference.

MWA relies on three types of events: *treatment* events, *dependent* events, and *control* events. The goal is to determine the effect that the occurrence of treatment events has on the occurrence of dependent events. To show actual causality, this effect is compared to the effect that the occurrence of corresponding control events has on the occurrence of dependent events. In the example of EA, *aid projects excluding parts of the community* are the treatment events and *insurgent activities* are the dependent events. The control events need to be selected such that they are reasonably comparable to the treatment events but still structurally different. In EA, the control events are *more inclusive aid projects benefiting the whole community*, like improving a village's infrastructure. Taken together, the treatment and control events represent the *intervention* events.

Simply comparing the effect of the treatment and control events, as depicted in Figure 1(S1), would not constitute a convincing causal inference, yet. Instead, it needs to be ensured that the events that are compared took place under conditions that are "as similar as possible." Otherwise, confounding factors could skew the results. For instance, if the control events took place in areas that are generally more affected by conflicts, it is also more likely for insurgent activities to occur in the same area, irrespective of the type of aid project conducted.

To ensure the similarity of the treatment and control events, MWA requires the *communicating researcher* (i.e., the expert whose application scenario of MWA is communicated) to provide so-called *matching variables* describing the general conditions under which the events took place. For instance, the population density of the area may play a role, or which ethnic groups live there. MWA performs *statistical matching* (Iacus, King & Porro, 2012) to only compare events with each other that are similar with respect to all provided matching variables. A corresponding visu-alization from our story is depicted in Figure 1(S3). One special matching variable is the *trend*. It is calculated by MWA directly and does not have to be provided by the communicating researcher. The trend describes the dynamic of the occurrence of the dependent events prior to treatment/control events. In the example of EA, it captures to which degree the number of insurgent activities were already increasing or decreasing prior to the start of the aid project, see Figure 1(S2).

The last core mechanic of MWA deals with the question of how large the area around the intervention events should be when counting dependent events in their proximity. This refers to both the spatial and temporal distance. The resulting effects may vary if, e.g., only five days before and after the start of an aid project are considered instead of 30 days when counting the surrounding insurgent activities. To solve this issue, MWA performs the analysis multiple times using different combinations of spatial and temporal distances. The results from the different analyses can then be compared, see Figure 1(S4).

Our story walks the user through all these mechanics. Like MWA, the story can be applied for different data sets and scenarios that use MWA, adjusting automatically. A communicating researcher only needs to provide their data, and our code used for the story automatically adjusts the views, creates appropriate textual interpretations of the views and results, and searches for the best example events in the data set to explain the different mechanics of MWA. Three such examples are provided on this website: https://matchedwake.com/. While MWA was originally introduced in the field of conflict research, it can also be applied to comparable problems in other domains, along with our story.

## 4 Designing an expert exploranation

In this section, we present our insights into how ExEx can be performed. The insights are based on our experience from designing the story to communicate MWA. We first discuss the goals and challenges of ExEx before elaborating on the design process we distilled from our story creation. Lastly, we explain what kind of visualizations and interaction techniques we used in the final story. Our audience ranges from graduate students to senior scientists in the field of conflict research. Their numeracy can be expected to be advanced, and as researchers with a statistical background, they can be expected to be familiar with basic visualization and interaction techniques.

## 4.1   Goals and challenges

**Goals.** Our goals, determined with the domain expert, were to provide researchers an *accessible* and *engaging* way to *understand* the core workings of MWA and allow them to judge whether the method is *applicable* in their own research. Moreover, the story should be *adaptable* to different application scenarios of the method. This allows the integration of the story not only along with the original method it explains but also in the context of research projects that used the method for their own research, e.g., by including it on the project website. There, it can allow anyone interested in the research to understand more easily how the method was used in its context. This way, also the scientific outreach of the original method can be increased.

**Challenges.** ExEx comes with certain challenges that go beyond the typical challenges of exploranation for broad audiences. The jargon of domain experts is more specific than the "lay language" of a broad public (Meredith, 2021). This makes the story designers more dependent on the experts' input when producing text for the story. Another point that requires additional input is to understand which domain-specific concepts need to be explained and, in contrast, what prior knowledge the users, who are other researchers from the field, can be expected to have. However, among experts, the prior knowledge can also vary, depending on how familiar they are with the concepts explained in the context of the method. To account for this, users should be provided with the option to skip parts that they already know.

Moreover, when using ExEx for explaining a method, interaction is not primarily used for exploring data sets, as it tends to be the case when addressing broad audiences. Rather, it is also used to better convey certain mechanics and the implications they have for the workings of the method. We provide examples for this later in Section 4.3.

Another point to consider is that designing visualizations and arranging them in a sequence that effectively conveys information becomes more challenging the more complex the conveyed information is. In addition, it requires more effort for visualization designers to understand a method from another research domain deeply enough to explain it to other researchers from that domain, as compared to explaining it to a broad audience on a much higher level.

Lastly, if the method to be explained is available as a software package, as much access as possible to the underlying code is desirable.

Overcoming these challenges requires close cooperation with domain expert partners. The increased scientific outreach and benefits for the research community that such stories can have can be valuable enough for the partners to contribute the time necessary (Beyer et al., 2020).

## 4.2 Steps of the story design process

In this section, we describe the steps we followed to design our story. We derived them by merging existing approaches (Lee et al., 2015; Cortes Arevalo et al., 2020; Zhang et al., 2022; Satyanarayan & Heer, 2014; Amini et al., 2015) to design data stories and adjusting them to the requirements of ExEx settings. The relationship between the corresponding works and our approach is shown in Table 1.

The process of cooperating with our project partner was inspired by Lloyd and Dykes's analysis of approaches for human-centered geovisualization design (Lloyd & Dykes, 2011). A central part played a Miro board,[3] which we will refer to as our "storyboard." Of course, the design process was not as separated and streamlined as conveyed through the sectioned structure below, but rather required several iterations.

### 4.2.1 Understand and abstract the method

The first step for the designer of an ExEx story should be to familiarize themselves with the method to be explained. When designing stories for a broad audience, the corresponding steps would be to *explore and analyze* (Lee et al., 2015) the data or to *read, interpret and select it* (Amini et al., 2015), see Table 1.

**Understand the method.** To understand the method, we referred to the corresponding paper (Schutte & Donnay, 2014) and discussed uncertainties with our project partner. This also required reading and discussing additional literature on which the method builds. While the underlying paper represented the most important source of information, for certain detailed questions, it was even necessary for our domain expert to dig into their code. Sometimes they even had to adjust it to let the corresponding package (Donnay & Schutte, 2023) return the necessary information, highlighting the importance of having access to the code of the communicated method.

---

[3]https://miro.com; accessed 23 October 2023

| Expert exploration process | | Lee et al. (2015) | Arevalo et al. (2020) | Zhang et al. (2022) | Satyanarayan and Heer (2014) | Amini et al. (2015) |
|---|---|---|---|---|---|---|
| **Understand and abstract the method** | 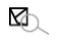 Understand the method | Explore and analyze | Prepare concept | Plan the message | Exploration to uncover interesting stories in data sets | Reading and interpreting data |
| | 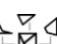 Abstract the key mechanics into story pieces | | | | | Selecting data |
| **Arrange the story pieces** | 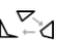 Determine the interdependency of the story pieces | Make logical connection | Define parts | Compose the information units | Drafting to prototype ways of communic. the stories found | Crafting the narrative structure |
| | 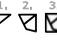 Arrange the pieces in an educationally meaningful order | Order story pieces | | | | |
| **Implement the story** | 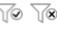 Determine the actor constraints | Build presentation | | Map the composition into visual | Production to develop the final interactive | Integrating strategies to engage viewers |
| | 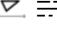 Draft text and visuals | | Draft text and visuals | | | |
| **Review and iterate** | 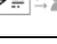 Review and iterate | Respond to input | Review and iterate | *Not explicitly mentioned* | | Non-linear and iterative process |

Table 1: The table shows the relation between the steps of expert exploration and other visual data story design process abstractions.

**Abstract the key mechanics into story pieces.** To keep the story at a reasonable length, not all details of its working and corresponding parameters could be explained. The abstraction process was based on iterative discussions and refinements using the storyboard. It resulted in a set of *story pieces* (Lee et al., 2015). This includes the main concepts and mechanics, like comparing different groups of events, i.e., treatment and control, and how the similarity of the compared events can be ensured via statistical matching. As expressed in Section 3, the mechanic of matching again requires other concepts to be explained, like the *trend* and how it is calculated.

### 4.2.2   Arrange the story pieces

In this step, the story pieces are arranged in an educationally meaningful order, similar to the steps *ordering the story pieces* (Lee et al., 2015) or *composing the information units* (Zhang et al., 2022).

**Determine the interdependency of the story pieces.** To do so, first, the logical relationship between the story pieces needs to be determined. Which concepts should be explained first? Which mechanics are based on which other? Flowchart sketches allowed us to discuss these questions. For instance, statistical matching depends, among others, on the trend. And to explain the trend, first, the concept behind treatment, control and dependent events needs to be explained.

This step is also important for identifying which aspects can be left out of the story. Overall, a trade-off has to be made between the breadth and the depth of the conveyed information. We aimed at keeping the story at around twenty minutes reading time. By analyzing the interdependency of the story pieces, it can be determined which of them are not relevant for understanding the main mechanics. If the method depends on input parameters and some of them are optional, this is also a good indication that the associated mechanics might not be essential.

**Arrange the pieces in an educationally meaningful order.** The determined dependencies between the story pieces help to bring them into an order in which they can be understood as easily as possible. In addition, user engagement and understanding can be increased by structuring the story along a story arc. For ExEx, the Freytag's Pyramid arc for visual data stories (Yang et al., 2021) can be adapted in the following way, see also Figure 1, in which the color coding corresponds to the following paragraphs.

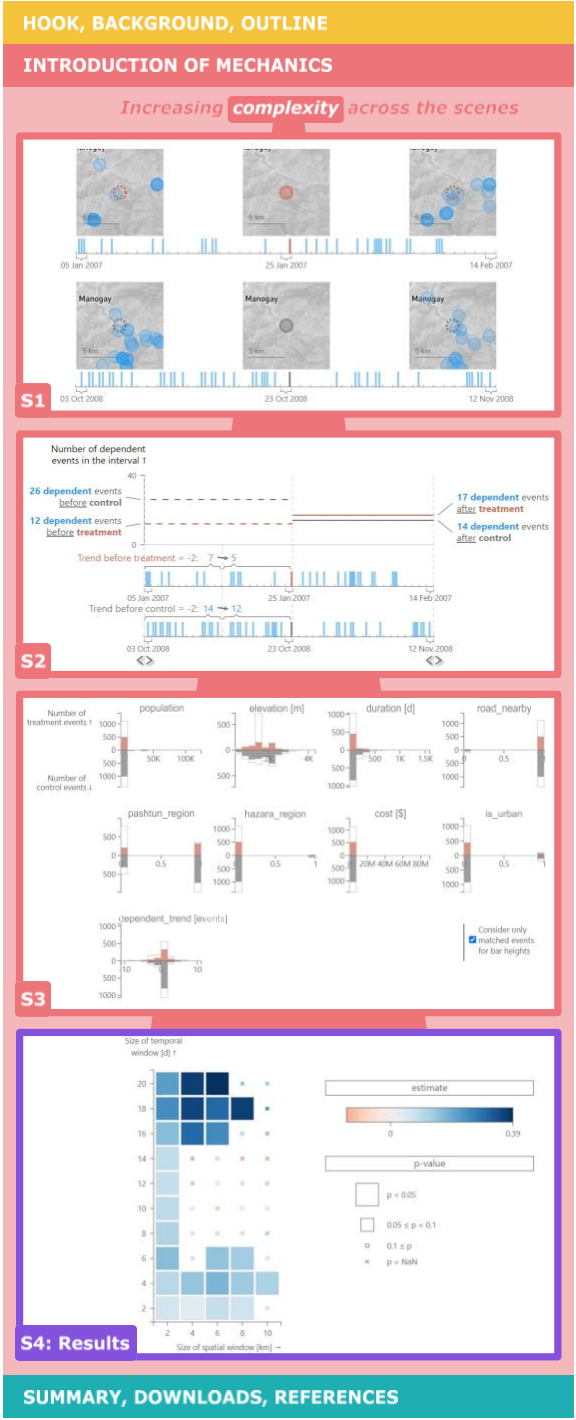

Figure 1: The figure shows the sections of our story, with the final shot of each core scene (S1-S4) represented by a screenshot. In S3, the variables *road_nearby*, *pashtun_region*, *hazara_region* and *is_urban* are binary.

First, the *setting* is established (Yang et al., 2021). In ExEx, this can include an initial question to **hook** the users, e.g., in case of the EA scenario: "What is the impact of aid projects excluding parts of a community on insurgent activities in Afghanistan?" Similar to applications for broad audiences, the hook is used to generate curiosity, e.g., by showcasing what kind of scientific questions can be answered using the explained method. Also, **background** can be provided regarding the purpose of the method and an initial **outline** of how it works. We presented all these three aspects in an introductory pop-up window.

In the subsequent core part of the story, *rising tension* leads up to a *climax* (Yang et al., 2021). In ExEx, this means that the key **mechanics** of the method are introduced step by step, building up to a comprehensive picture. In the *climax*, the **results** are revealed, including the answer to the initial question from the **hook**.

Lastly, in the *resolution*, the story is concluded (Yang et al., 2021). In ExEx, this can be a brief **summary** of the method, options to **download** the results, and a list of **references**. In addition to the overall arc across the entire story, individual sections of it may have peaks on their own. Amini et al. noted this as beneficial if multiple aspects of a topic need to be explained (Amini et al., 2015). To describe how we implemented this, we use the following terminology: The core of our story consists of four main sections, represented by the four screenshots in Figure 1. We refer to each such section as a *scene*. Each scene consists of a discrete set of views, which we refer to as *shots*.

In each scene, a key mechanic is explained, following the same miniature arc in all four scenes: Every scene begins with a relatively plain view. Then, information is continuously added with each shot until the view reaches its peak complexity in the final shot, as sketched in Figure 2 (D). This way, we aimed at making the explanations as easy to understand as possible (Amini et al., 2018).

With this *staging*, we also tried to keep the story consistent, even from one scene to the next, to maintain *continuity* (Roth, 2021). Such interconnectivity can make stories easier to follow and memorize (Zhang et al., 2022). Moreover, the final shot of each scene allows the user to interact with the visualization to engage more deeply with the information conveyed in the scene, deepening their understanding.

Following the approach of *exploranation* (Ynnerman, Löwgren & Tibell, 2018), these *interleaved explorative microenvironments* are designed to be expressive while also constraining the interaction to limit the risk of producing views that are diffi-cult to understand. With this approach, we also tried to account for the cycles of engagement and disengagement described by O'Brien and Toms (O'Brien & Toms, 2008). Reengagement with an application occurs when the user invests themselves in the interaction beyond a routine level. Hence, the interactive shots at the end of each scene encourage the user to reengage with the content and reflect before transitioning to the next scene.

For planning the arrangement of the story pieces, we used basic flowcharts in our storyboard.

### 4.2.3 Implement the story

When implementing a story, thorough storyboarding and prototyping can help reduce the need for time-consuming re-iterations at later stages (Hlawitschka et al., 2020). For storyboarding, we referred to our Miro board. While it was primarily used for flowcharts in the previous steps, for this step, we used the components depicted in Figure 2.

For each scene, the board contained an area for the following components: (A) A textual summary of what should be explained and visualized in the scene, in terms understandable to the domain expert. (B) A list of constraints and considerations for the scene, e.g., regarding important points to mention, potential special cases, visuals, and interaction. (C) A sketchboard area for designing and refining the different components of the visualizations, and (D) an area where the carved-out sequence of individual shots from the scene can be arranged. (E) In addition, the designers and the domain expert were able to place and reply to sticky notes wherever they needed to clarify something.

For prototypically implementing the views that were crafted in the storyboard, we used Observable to create a collection of cell-based online notebooks.[4] As the notebooks are web-based, they can easily be shared with cooperation partners. This allowed us to perform fast implementations and exchange feedback early on in the design process, as advised by Hlawitschka et al. (2020). Such exchange about the possibilities on the implementation level is key for close cooperation (Beyer et al., 2020). Observable provides easy inclusion of inputs like sliders to modify various aspects of a visualization and test border cases, which we made use of in our notebooks. This was beneficial as our story should be applicable to various data sets. The code from the notebooks also provided a solid foundation for implementing the actual story, which we did using React[5] and D3.[6]

---

[4]https://observablehq.com/collection/@zykel/exex-prototypes; accessed 24 October 2023

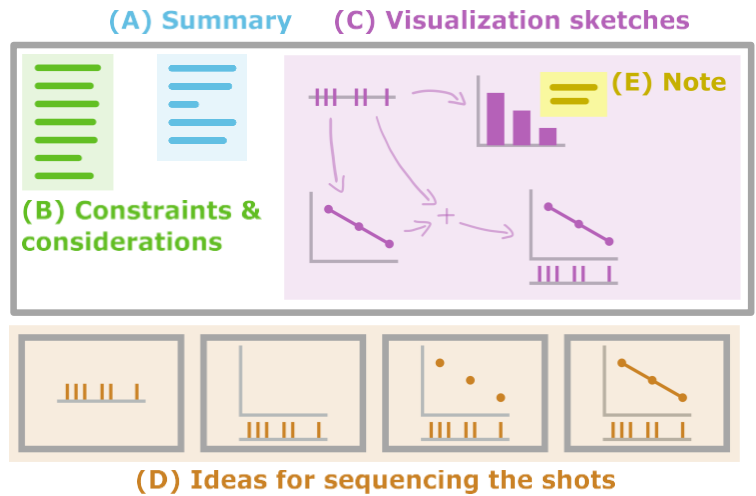

Figure 2: The figure exemplifies what the components of our storyboard looked like for a single scene. We used these components for each scene.

**Determine the actor constraints.** For better comprehension, we focused on selected example events from the underlying data set when explaining the different mechanics. These example events play the main roles in the story, so we refer to them as the *actors*. Drawing the actors from the data set instead of creating "ideal toy examples" makes the underlying data more tangible. Moreover, we wanted to keep the same actors in focus across the scenes. This way, it is easier to follow the story's main thread and grasp the interaction between the different mechanics. However, when drawing examples from the actual data set, it is not trivial to ensure that the examples are suitable to explain all mechanics equally well. Therefore, we thoroughly examined the story pieces to determine the constraints that the actors should fulfill. We weighted the constraints to be able to automatically determine the actors in arbitrary data sets.

**Draft text and visuals.** As visualization researchers, this step was the most extensive. We group our remarks based on the topics *genre*, *text*, *text-vis combination*, *staged animations*, *visual style*, and *flow factors*.

---

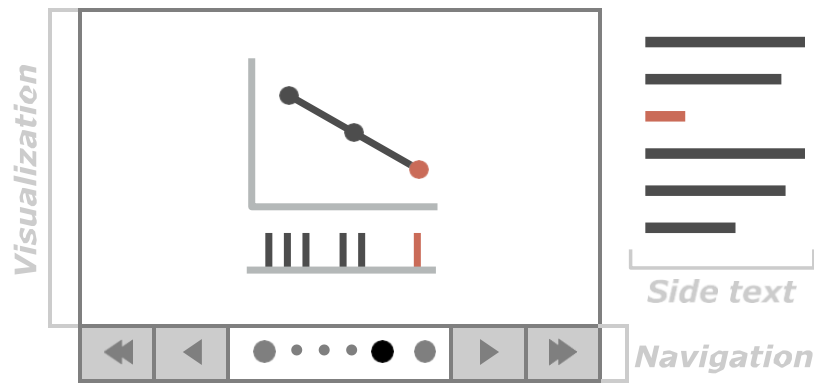

Figure 3: A single shot of our story consisted of a visualization and navigation component on the left and a text component on the right. The red color exemplifies a linking between visuals and text.

*Genre.* We created the story as a *dynamic slideshow* (Roth, 2021) with the corresponding slideshow-style layout, see Figure 3. Slideshows provide control over the story pacing as they limit users' skimming and allow to deliver key plot points more clearly (Roth, 2021). In addition, they allow to keep content persistent across slides to incrementally build up more complex views (Roth, 2021), and they provide a clear mapping between text and visualization (Zhi, Ottley & Metoyer, 2019).

*Text.* We drafted text at different stages of the design process. Initially, we outlined the scene contents and collected key points on the storyboard. During the development, we roughly decided what text should be provided in unison with the visuals. In the end, we finalized the text drafts with our domain expert, making sure to follow data provenance guidelines like providing data sources and additional references (Hullman & Diakopoulos, 2011). As the story should be adaptable to different data sets, certain parts of the text had to be kept variable, including automatic interpretations of results.

*Text-vis combination.* A dedicated visualization panel took up the larger fraction of the screen space, with a smaller text panel accompanying it to its right, see Figure 3. This way, the visualizations received the main focus to "foreground the topic through perceptual layering" (Ynnerman, Löwgren & Tibell, 2018). By revealing each new view in synchrony with the corresponding text, their connection was established, while linking individual textual and visual elements via color (Mayer et al., 2023) like in Figure 3. In general, we often verbalized what was depicted in the visualizations and stressed visual differences in the text, e.g., between certain groups of data points (Latif, Chen & Beck, 2021). The most common types of annotations we used in the visualizations were "text," "shapes" like arrows, and "highlights" of important areas (Ren et al., 2017).

*Staged animations.* When building up each scene into a final view, most of the visual changes in our story are animated. This can lead to less attention drift and better comprehension for the users (Amini et al., 2018) and increase their engagement (McKenna et al., 2017). We also break up the animations into multiple transitions between the individual shots. Such staging was shown to improve recall and understanding (Heer & Robertson, 2007). Within each scene, we tried to keep the transition costs minimal, i.e., the number of visual changes when going from one visualization to the next (Hullman et al., 2013).

*Visual style.* Despite the positive effects of embellishments on memorability (Bateman et al., 2010), for ExEx, they should be used thoughtfully to maintain a professional and credible appearance. However, a visually attractive style can let users have more patience and look at visualizations more deeply (Cawthon & Moere, 2007). We aimed for a clean and aesthetically pleasing style but, overall, the preference regarding the visual style may vary depending on the targeted research community. In addition, there is the question of how closely the visualizations in the story should stick to the style and content of the figures used in the underlying paper. High similarity makes it easier to connect the information from the story and the paper. However, it may result in missed opportunities for improving upon the design from the paper, including non-static options like interaction and animation. We took inspiration from the paper but reworked two visualizations (Scenes 1 and 4) and created two entirely new visualizations (Scenes 2 and 3).

*Flow factors.* Flow factors capture how the navigation of a story interacts with its visual components (McKenna et al., 2017). As navigation control, we provided buttons. They allow the user to either transition to the next shot with an animation or to skip the animation and fast-forward immediately to the next shot, see Figure 3. The navigation progress could be tracked via so-called *breadcrumbs*, which could also be used for navigation, see Figure 3. The story progression was intended to be linear, going from one shot to the next, but the user also has the option to skip ahead using the breadcrumbs if they are already familiar with certain concepts and want to save time. Overall, our story would be classified as a *stepper*, which was the most engaging story style in the experiments of McKenna et al. (2017).

#### 4.2.4 Review and iterate

Reviewing and iterating on our story was particularly relevant for ensuring that the designers had correctly understood and explained the relevant mechanics of the communicated method. Having to communicate such complex information, we did not decouple the implementation of the visual aspects entirely from the narrative structure, contrary to other approaches (Satyanarayan & Heer, 2014). Creating preliminary visualizations was necessary as a basis for discussion. However, depending on how big the resulting changes are, they can be quite impactful on the story, since the explanations build on each other, as do the visual design decisions. Therefore, good abstractions and storyboarding processes are still important to keep the later changes at a minimum. For storyboarding, Amini et al. observed a highly iterative and non-linear process (Amini et al., 2015). In their study, participants often alternated between *reading and interpreting data*, *selecting data*, *crafting the narrative structure*, and *integrating strategies to engage viewers*. This was also the case in our design process, for which our storyboard and prototyping notebooks were of great help.

Note: Some of the approaches from Table 1 also include the evaluation of the story as part of a storytelling design process (Lee et al., 2015; Cortes Arevalo et al., 2020). We describe our evaluation separately in the next section, as it is more extensive than what will be possible under the circumstances commonly present when performing ExEx. The reason for this is that we could refer to two complementary user groups which will not always be possible.

### 4.3 Visualization design

We give an overview of the central visualization design decisions and how they supported the explanation of the method. Each paragraph corresponds to one core scene, see Figure 1.

**Scene 1.** Three map views are displayed for an example treatment event, and three maps for a matching control event. The maps to the left display the dependent events before the intervention, and the maps to the right the dependent events after the intervention, each with a dashed circle as a reference to the location of the intervention event. This way, the different time intervals are separated and the treatment and control scenario are contrasted. In addition, timelines represent the dates of the events. Synchronized panning and zooming of the maps corresponding to the same scenario allow closer examination of the events, combined with additional information upon hovering.

**Scene 2.** We display the aggregated counts of the dependent events before and after the interventions. Visual aids above the timelines explain how the trend is calculated. Annotations highlight the key information and adjust upon interaction. To interact, users can change the size of the temporal window by dragging the handles at the bottom of the view to see how the counts and the trends would change if a different temporal window was used. If the trends of the dependent events prior to the interventions become too dissimilar, visual highlights reflect that, now, the intervention events would be considered too dissimilar to constitute a valid pair for comparison.

**Scene 3.** One pair of histograms is displayed for each *matching variable*. The matching variables are additional data provided separately to cover for potential confounding factors. They are provided by the communicating researcher. The histograms provide an overview of how the treatment and control events are distributed across the matching variables. This is relevant because events can only be *matched*, i.e., included as valid pairs for performing inference, if they are similar enough across all matching variables. The views can be toggled to only display the matched events, and on each pair of histograms, a filter can be applied to remove the events across all histograms that do not fulfill the filter constraint. This way, users can get a better impression of how the data is distributed, and it also becomes clearer how the likelihood of events being matched is reduced the more matching variables the communicating researcher includes. For applying MWA, ideally, the treatment and control distribution are very similar for each matching variable. The more dissimilar the distributions are, the more dissimilar are the underlying events, meaning they are less likely to be matched.

Note that the x-axes are not annotated directly, only via the respective chart titles which correspond to the names of the matching variables in the data set. Accordingly, no separate information about the units is provided if the communicating researcher does not include it in the variable names. This solution was preferred by our collaborating expert as it keeps the additional information minimal that a communicating researcher has to manually provide when creating a new version of the story based on their own research problem solved with MWA.

**Scene 4.** A heatmap gives an overview of the results of the analysis when performed for different spatial and temporal windows. The effect size estimate is mapped to the color and the significance of the results to the size of the tiles. This allows to quickly detect patterns in the results and to assess for which window sizes the effects are meaningful. Hovering over a tile yields the precise effect size and highlights it in the legend.

# 5 Evaluation

For the evaluation, we invited eight peace and conflict researchers. Four of them had used MWA for at least one of their papers before, the other four had not. Hereafter, we refer to the first group as *experts* and to the second group as *semi-experts*. The semi-experts were two Ph.D. students and two post-doctoral researchers. The experts all had a Ph.D. and between three and 13 years experience in conflict research. All participants see data visualizations on a regular basis, with all but two of the experts producing visualizations themselves.

For each expert, we created a version of our story based on one of the papers in which they had applied MWA. This resulted in three stories, as two experts had worked on the same paper. With each participant, we scheduled an interview in which, after a brief introduction, they were given the link to one version of the story. For the experts, we used the story corresponding to their paper, and for the semi-experts, we used the story based on the example paper (Karell & Schutte, 2018) used in Section 3. While following the story, the participants shared their screen and could ask questions or give feedback. Afterwards, we collected additional verbal feedback and sent them a questionnaire to fill out later. The interviews lasted 45 min on average (38 min median) of which reading the story took 23 min on average (20 min median). In the evaluation, we wanted to validate two main aspects:

(A) How engaging and how understandable do the participants find the story and its content?

(B) What do the participants think about the approach of ExEx in general?

Note: While memorability is also an important aspect in visual storytelling, we did not focus on it, as it was not central to our overall narrative intent of explaining the method.

**Regarding (A):** We were particularly interested in the participants' engagement and the understanding, as well as how beneficial the exploratory views were for them. To test the engagement, we referred to the 5-point Likert scale questions proposed by Hung (2019). As Hung focused on measuring engagement for individual visualizations and we wanted to test an entire story, we replaced the occurrences of "visualization" in the items by "tutorial." The resulting items are listed in Table 2 (E1-E11).

To test how well the participants understood the content of the story, we adapted the approach by Burns et al. (2020). They used the *taxonomy of educational objectives* originally proposed by Bloom (1956) to evaluate six different dimensions of understanding in data visualization. For our work, the dimension of *comprehension* was the most central. It refers to how well "learners understand the underlying information as a whole" (Bloom, 1956). Burns et al. evaluate this by asking the participants in the open-ended question how they would summarize the conveyed information to explain it to a friend. We followed this example and asked the participants the following question after they had finished the story: *"Imagine that you want to explain the content of the tutorial to a peer without showing them the tutorial. How would you describe the method explained here (Matched Wake Analysis) in your own words?"* (U0)

The participants were instructed to answer the question verbally. Aside from *comprehension*, we were also interested in the dimensions *knowledge* and *application* (Burns et al., 2020). *Knowledge* refers to whether a user is able to "recall or recognize factual information" (Bloom, 1956). *Application* describes whether a user is able to "break down a topic into parts and understand the relationship between each part" (Bloom, 1956). We covered these two dimensions in our analysis of the answers to question U0 as it also depends on them.

In addition, we wanted to know whether certain aspects in the story were unclear or explained in too little or too much detail. Accordingly, for each of the four main scenes of the story, we included items U1-U6 from Table 2. For each scene, we also included items I1-I4 to test how well the interactive components were perceived.

**Regarding (B):** To investigate the participant's general interest in expert exploranation, we included items G1-G6 from Table 2. In addition, we collected verbal feedback during the interview.

| | ID | Question |
|---|---|---|
| **Engagement** | E1 | My use of the tutorial is continuous and smooth. |
| | E2 | I feel motivated while using the tutorial. |
| | E3 | I DON'T feel frustrated when using the tutorial. |
| | E4 | I enjoy exploring the tutorial. |
| | E5 | I have acquired a new concept or new knowledge from the tutorial. |
| | E6 | I think the tutorial effectively delivers its main concept or idea. |
| | E7 | I think the tutorial is telling a compelling story. |
| | E8 | I think the tutorial sparks my creative thinking. |
| | E9 | I feel entertained when using the tutorial. |
| | E10 | I feel absorbed by the tutorial while using it. |
| | E11 | The look and feel of the tutorial is pleasing to me. |
| **Understanding (f.e.s.)** | U1 | I think there are unclear concepts or explanations in this section. |
| | U2 | I think the following concepts or explanations are unclear. *(Free text answer)* |
| | U3 | I think there are aspects that should be explained in more detail. |
| | U4 | I think the following aspects should be explained in more detail. *(Free text answer)* |
| | U5 | I think there are explanations that go into too much detail. |
| | U6 | I think the following explanations go into too much detail. *(Free text answer)* |
| **Interaction (f.e.s.)** | I1 | I found the interactions appropriate. |
| | I2 | I found the following interaction options unnecessary. *(Multiple choice)* |
| | I3 | I would have liked to explore certain aspects in more detail. |
| | I4 | I would have liked to explore the following aspects in more detail. *(Free text answer)* |
| **General** | G1 | I would recommend the tutorial to a fellow researcher who wants to familiarize themselves with MWA. |
| | G2 | I would use a tutorial like this instead of reading the underlying paper. |
| | G3 | I would use a tutorial like this in addition to reading the underlying paper. |
| | G4 | I think that the tutorial can help researchers assess whether MWA is applicable to their research questions. |
| | G5 | I think the following information is missing from the tutorial to help researchers assess whether MWA is applicable to their research questions. *(Free text answer)* |
| | G6 | In my research, I see the following other applications where this kind of tutorial can be beneficial. *(Free text answer)* |

Table 2: The questionnaire consists of 5-point Likert scale questions (aside from certain exceptions declared in brackets). The blocks Understanding and Interaction were asked for each scene (f.e.s.). Questions G4 and G5 were formulated in first person for the semi-experts.

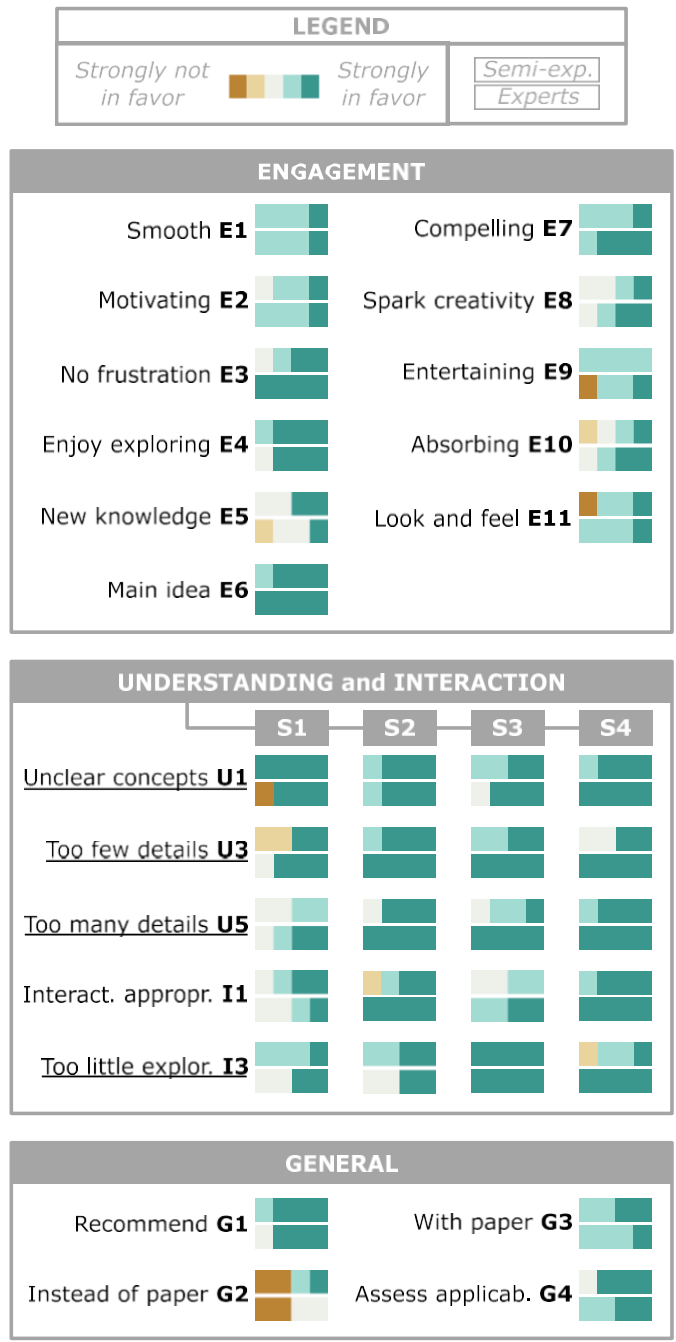

Figure 4: The figure shows the ratings from the Likert scale questions. The bars for the ratings of the semi-experts are always depicted above the bars for the experts' ratings. The bold item IDs correspond to the IDs in Table 2. For negative items (which we underlined), we have flipped the color scale, so green corresponds to answers in favor of our stories across all items. The abbreviations "S1" to "S4" stand for "Scene 1" to "Scene 4."

## 5.1 Results of the evaluation

In this section, we use the item IDs depicted in Table 2 and Figure 4 when referring to specific items from the questionnaire. Overall, the participants received the stories very well. They liked the direction of conveying methodological understanding in this way and mentioned that the "narrative aspect of it helps to contextualize [the presented information]." They also found the visualizations, animations, and interactive elements "extremely helpful" and "very valuable" as they "create an intuition that just cannot be created verbally." They said that the direction was "incredibly important" but "underexploited" and that "the future [lies] here." In the following, we group the quantitative and qualitative feedback based on overall topics. We primarily discuss the critical feedback, though the participants expressed that their complaints were already at a high standard.

### 5.1.1 Engagement.

As depicted in Figure 4, the overall engagement was quite high, with the following exceptions.

*New knowledge and concepts.* The question whether new knowledge and concepts were acquired was rated mixed, as the experts already knew MWA and two of the semi-experts had already read the paper before. However, participants who had already read the paper stated that the story helped them to recall information that, for some, was multiple years old. In addition, for those semi-experts who had not read it before, the learnings were large, allowing them to summarize the workings of MWA in surprising depth after using the story. Remarkably, also one of the experts gained new insights, as Scene 3 made them think about the mechanics of MWA in a way they "didn't [before], because we never made that graph."

*Entertainment.* The rating of one expert showed that they were not entertained during the story. This expert also remarked that the pacing of the story was generally too slow for them as they were already familiar with a lot of MWA's underlying mechanics. They suggested that a more time-efficient version of the story would suit them better. Other participants suggested that, at the beginning of each scene, a high-level summary would help, stating which scientific problem is solved using which approach in that scene. For instance, in Scene 3, a "potential selection bias" is prevented using "statistical matching." This way, users already familiar with the underlying approach of statistical matching could skip ahead and save time.

*Look and feel.* A semi-expert found the aesthetics of the story unpleasing (E11). They expressed that they strongly disliked the red color used for treatment events and would have preferred a monochromatic color scale across the entire application. In contrast, an expert said that the choice of colors was good and "made absolute sense." In general, the participants perceived the look and feel of the story very well, describing it as "amazing," "very cool," and "very good." They also used all navigation options, going forward and backward with and without animation options, using the breadcrumbs to jump to specific shots, and even following links to external websites with further information. However, the linear navigation through the story using animation was the default for all participants, showing that the animations were relevant for them. In addition, the decision to change and add new visualizations compared to the figures from the underlying paper was appreciated. One expert even stated that when they had worked with MWA, they "used the default output visualization of MWA [which was also used in the original MWA paper], but yours [in Scene 4] is much more clear." Moreover, the fact that the visualizations in Scene 2 and 3, which we had created entirely new, led to new insights for some experts, also showed the benefits of going beyond the figures used in the original paper.

### 5.1.2    Understanding and Interaction.

Regarding the request to summarize MWA after using the story (U0), the participants answered quite similarly. Overall, the key points related closely to the four main mechanics presented in the four different scenes of the story. The relationships between the different parts were reproduced well, speaking for good *application* (Bloom, 1956). Regarding the *knowledge* dimension (Bloom, 1956), all recalled information in the summaries was factual. The key difference was that some participants were more explicit about certain aspects than others. Most remarkably, one semi-expert and one expert did not explicitly mention that there are two different types of intervention events, the treatment and control events. Upon request, the participants who were already familiar with the method said that it substantially helped them to refresh the memory of the method, and that they could have not summarized it in such detail anymore without using the story. Overall, the *comprehension* (Bloom, 1956) seemed very good.

This is also reflected in the self-reported understanding from the quantitative feedback. It was very positive, with a few exceptions that we discuss in the following. Overall, Figure 4 conveys that most understanding issues occurred in Scene 1. However, the corresponding free text feedback revealed that the remarks rather required more context regarding the overall story instead of criticizing Scene 1, specifically.

*Context.* Contrary to the point mentioned before about improved time-efficiency for experts with strong prior knowledge, it was also remarked that, for a certain part of the scientific audience, the prior knowledge can be relatively low. Conflict research is special in that regard, as there is a divide between qualitative and quantitative scholars. According to one participant, to allow also experts with little quantitative background to follow the story, aspects like the basic idea of *scientific comparison* between treatment and control groups would need to be explained. Accordingly, in the beginning, the big picture could be summarized in a more abstract way, e.g., that the intuition behind the approach is that "in the data, there are some natural experiments hidden that have to be found, and the right things need to be compared for doing so." This could "buy the attention" of a less quantitatively-oriented audience.

In addition, a participant requested whether the final textual interpretation of the results could be made even more clear, regarding which units the calculated effect size is represented in.

*Text structure.* It is not visible from Figure 4, but in the third scene, some participants had minor issues following the main thread of the scene. It was caused by some of the text blocks being quite large, with too little visual structure. Three participants suggested adding more paragraph breaks or using a bullet point text structure. One participant also asked whether the text could be embedded more directly into the visualizations.

*Interaction.* Overall, the provided interaction options were perceived very well. They were explored by all participants, most deeply the interactions in Scene 2 and 4. Some participants did not see the necessity to pan and zoom maps in Scene 1 and to draw filter constraints into bar charts in Scene 3. The more straight-forward interaction options were adopted better, i.e., hovering over elements for additional information (Scene 1 and 4) and dragging a handle (Scene 2). However, despite their ease of use, these options yielded substantial additional information. In contrast, the option to (un)tick a checkbox in Scene 3, which did not reveal entirely new information, was seen as unnecessary by three participants.

### 5.1.3 Applicability

All participants expressed the desire to use a story such as ours in addition to the underlying paper for familiarizing themselves with a new method (G3). However, the preferences regarding when to use it (before, while, or after reading the paper) differed. All were mentioned in the oral feedback, but, primarily, the participants would use it while reading the paper, particularly when reading the method section. However, a difficulty mentioned was that, to do so, people would need to be made aware that the story exists before reading the paper. One way would be to make sure that it can be found when looking up the method on search engines.

As expected, in most cases, the participants would not use the story as a replacement for reading the paper (G2). They would only do this if they wanted to get an overview of multiple existing methods, but not if they actually wanted to apply the underlying method in their own research. They expressed that the reason was not an issue of trust, but the fact that the paper contains more details than the story.

Overall, though, the participants thought that the story can help researchers assess whether MWA is applicable to their research (G4) and that they would recommend it to fellow researchers (G1). All participants expressed a strong desire to use a tailored version of the story to explain to other researchers how MWA was used in their own research. One expert even asked whether they could share their version of the story on X (formerly Twitter).

Aside from MWA, the participants immediately came up with other methods that would benefit from ExEx, including complex analytical methods in general, like causal and spatial analysis, but also concrete methods (Kelly, 2019). They also saw great potential for ExEx stories in teaching, primarily for students in their Master's and onward. In addition, it was mentioned that such stories can help less technical project partners understand how a complex method was used in a collaborative project.

Adding to that, one expert said that ExEx could even help those researchers who have already understood and applied the explained method. Accordingly, the expert who received a new perspective on MWA when they used our story said that such a story "can challenge you on how much you really understood - that really happened to me." Moreover, two experts said that such a story can be helpful for reviewers when submitting a paper that used the method explained in the story as "you can send the link with it like: here, in a nutshell."

For such cases, when the story is used by experts with a strong background, and time efficiency plays a bigger role, again, the approach mentioned before could be beneficial: Equipping the story with a technical summary at the beginning of each scene would help to allow users already familiar with the explained mechanic to skip ahead and save time. In this case, though, the users should be able to select whether they want this more advanced version of the story or just the standard version. This would help to prevent the technical summaries from scaring away users with less prior knowledge.

Lastly, one participant mentioned that it could also be nice to have the option to export animated data-GIFs (Shu et al., 2021) from certain transitions, or to export individual (static) views, including the option to load new data into the view.

# 6   Reflections

We summarize the insights obtained from our evaluation and the previous works we used as the foundation for our design process. In that, we refer to the participants of the evaluation simply as "participants," and we do not list all the references to the previous works again that were already listed in Section 4. Of course, our suggestions should not be taken as strict rules. Further investigations are necessary to validate the applicability of the suggestions on a broader set of scenarios. Moreover, it would be beneficial to evaluate certain aspects more carefully, like how the jargon used in the story was perceived, and which of the concepts explained in the story were already familiar to the users.

*Don't assume knowledge from the paper.* The feedback showed that potential users may want to use an ExEx story while or even before reading the corresponding paper. Therefore, no prior knowledge from the paper should be assumed in the story. Moreover, the story does not have to replace the paper, as most participants expressed that they would rather use it to complement the paper which is more detailed than the story.

*Cooperate closely.* As in other visualization projects, close cooperation with the domain experts is key. In ExEx, it is relevant for making sure that the story designers have properly understood the underlying method, to identify which mechanics are most relevant to explain, and for drafting the final story text. Moreover, as much access as possible to the implementation of the explained method is beneficial, and iterations will most likely be necessary.

*Prototype thoroughly.* When designing an ExEx story, it is important to prototype it thoroughly, as changing substantial aspects at a later stage can be very costly since the different parts of the story (including visual design decisions) build on each other. Abstracting the method well and determining good actor constraints at an early stage is key, as is the selection of efficient prototyping tools. For bridging the gap between prototyping and implementation, Observable served us well.

*Use an appropriate layout.* Based on the insights from previous works, we selected the slideshow genre for our story. It allows for a clear mapping between visualization and text, the possibility to progressively build up views, and straightforward navigation. We strengthened the mapping between text and visuals by linking them via color, annotations, and verbalizing the visual information. However, as suggested by one participant, interleaving text and visuals even more might reduce the effort for shifting the focus between visuals and text back and forth with each new shot. Moreover, the accompanying text should be structured well to avoid big blocks of text.

*Stage the scenes.* In each scene, we built up a visualization step-by-step, breaking down a complex mechanic and a correspondingly complex visualization to make it more accessible and easier to understand. Animating the transitions helps to make the visual changes easier to follow and was received well.

*Provide interactivity.* The participants adopted the interaction options well and gained new insights from them. The options that were straight-forward but at the same time yielded new information were perceived best.

*Dare to invent visualizations.* When creating visualizations, we found that sticking exactly to the figures used in the underlying paper was not necessary. We based some of our visualizations on figures from the paper, but also created new ones. This helped the participants to see certain insights more clearly or to even gain new insights they had not gained when working only with the paper and the underlying method.

*Provide clear interpretations.* While we already provided automatic interpretations, it showed that they could have been even more precise, stressing the requirement of having clear automatic interpretations of visualizations and results.

*Provide context and outline.* Both at the beginning of the story and at the beginning of each scene, a clear outline should be given of the information to be presented. Also, the context in which the explained method can be applied should be given.

Moreover, if the method is explained by following an example application, also the context of this example application should be provided. We tried to do so, but the participants would have liked even more such overview information. It was also confirmed that, even for an expert audience, the prior knowledge of the users can vary strongly. Accordingly, to not overwhelm users with limited prior knowledge, there should not be too many technical terms in such overviews. At the same time, to not lose the experts with a strong background, we suggest the next point.

*Consider a time-efficient version.* Particularly for experts with a strong background, it may pay off to provide a version of the story where the contents are summarized on a technical level at the beginning of each scene. This way, more knowledgeable experts would be able to progress through the story faster, only stepping through a scene in detail if they are unfamiliar with the explained concepts. The detailed version of the scene could be the same as for the original stories, so only little additional implementation effort would be required. At the beginning of a story, users could be given the option to choose whether to use the more advanced version or the original one. However, it needs to be noted that experts can gain new insights even if they are already familiar with a certain topic, as it was the case with one participant. Yet, for this to happen, they also need to be open and attentive enough when reading information about a topic they are already familiar with.

*Consider the benefits.* As with all visual storytelling applications, a considerable amount of work goes into the design and implementation of an ExEx story. This also includes the domain expert whose close cooperation is needed, see the earlier point. However, the broad applicability of such stories and the corresponding increase of outreach can help to motivate experts to put in the extra work necessary (Beyer et al., 2020). Accordingly, the participants already had a number of additional applications in mind for which the stories could be used, ranging from teaching to supporting analysts that are already using the explained method.

Following Meredith's line of argument, putting effort in the creation of such stories can pay off even more (Meredith, 2021). Accordingly, additional audiences aside from the intended research colleagues could be reached, like potential students or stakeholders, if the jargon and the level of detail are adjusted. A natural next step in this direction would be to conduct a study with students to investigate how strongly an ExEx story would need to be adjusted to support them in their education.

*Automate.* To reduce the work of the designer, research on automating the process of ExEx is crucial. While we have identified potential challenges and guidelines for ExEx in this work, more automated processes need to be built on them to make the creation of ExEx stories more accessible and economically reasonable on a larger scale. Current research, e.g., regarding more complex slideshows, is still lacking (Chen et al., 2023). A good first step for facilitating the story creation is to provide structured access to the byproducts from the creation of the content to be explained, i.e., in our case, the scientific method. Code-based notebooks are a great way to do so, as we also benefited strongly from the notebooks we had created in our ExEx process.

*Make it visible.* Informing potential users of a method about the existence of an accompanying ExEx story in time is key for supporting their familiarization with the method. The feedback showed that some users might use such a story even before reading the paper, and most while reading it. Providing a link to the story on the method's project website or in a corresponding package would be beneficial. Sharing it on social media can also help, as one participant offered to do. Ideally, the story could be mentioned already in the paper, though, at the point of submitting the original paper, such a story probably does not exist, yet.

# 7   Conclusion

We analyzed how exploranation can be applied to support the communication of scientific methods between researchers. We focused on explaining a method for spatio-temporal causal inference between conflict events by producing a story that can automatically adapt to specific application scenarios of the method. We abstracted our design process and evaluated different versions of our story with eight conflict research experts, of which four were already familiar with the method.

The results show that expert exploranation is a promising direction, as the experts received the story very well, seeing various applications for such stories. Their feedback allowed us to derive suggestions for how expert exploranation can be performed, ranging from the visual and textual design of the story to making it more efficient for users with differing background knowledge. Yet, substantial research is still necessary, both for validating our findings to produce more generalized recommendations, and for allowing automation in the creation of expert exploranation stories to make it more time-efficient.